\pdfoutput=1

\documentclass[pra,superscriptaddress,amsmath,amssymb,twocolumn,nofootinbib]{revtex4-2}

\usepackage{graphicx,bm,amsmath}
\usepackage[usenames,dvipsnames]{xcolor}
\usepackage[colorlinks,bookmarks=false,citecolor=NavyBlue,linkcolor=Red,urlcolor=blue,
]{hyperref}

\usepackage[bbgreekl]{mathbbol}
\usepackage{xcolor}
\usepackage[normalem]{ulem}
\usepackage{algorithm}
\usepackage{braket}
\usepackage{algpseudocode}
\usepackage{comment}
\usepackage{amsthm}
\usepackage{enumerate}

\def\doi{http://dx.doi.org/}

\newcommand{\be}{\begin{equation}}
\newcommand{\ee}{\end{equation}}
\newcommand{\bec}{\begin{equation*}}
\newcommand{\eec}{\end{equation*}}
\newcommand{\bea}{\begin{eqnarray}}
\newcommand{\eea}{\end{eqnarray}}

\newcommand{\titleinfo}{Magic transition in measurement-only circuits}

\newcommand{\Tr}{\text{Tr}}   
 
\begin{document}
\title{\titleinfo}
\author{Poetri Sonya Tarabunga}
\affiliation{International School for Advanced Studies (SISSA), Via Bonomea 265, I-34136 Trieste, Italy}
\affiliation{The Abdus Salam International Centre for Theoretical Physics (ICTP), Strada Costiera 11, 34151 Trieste, Italy}
\affiliation{INFN, Sezione di Trieste, Via Valerio 2, 34127 Trieste, Italy}

\author{Emanuele Tirrito}
\affiliation{The Abdus Salam International Centre for Theoretical Physics (ICTP), Strada Costiera 11, 34151 Trieste, Italy}
\affiliation{Pitaevskii BEC Center, CNR-INO and Dipartimento di Fisica,
Università di Trento, Via Sommarive 14, Trento, I-38123, Italy}

\begin{abstract}
Magic, also known as nonstabilizerness, quantifies the distance of a quantum state to the set of stabilizer states, and it serves as a necessary resource for potential quantum advantage over classical computing.
In this work, we study magic in a measurement-only quantum circuit with competing types of Clifford and non-Clifford measurements, where magic is injected through the non-Clifford measurements. This circuit can be mapped to a classical model that can be simulated efficiently, and the magic can be characterized using any magic measure that is additive for tensor product of single-qubit states. Leveraging this observation, we study the magic transition in this circuit in both one- and two-dimensional lattices using large-scale numerical simulations. Our results demonstrate the presence of a magic transition between two different phases with extensive magic scaling, separated by a critical point in which the mutual magic exhibits scaling behavior analogous to entanglement. We further show that these two distinct phases can be distinguished by the topological magic. In a different regime, with a vanishing rate of non-Clifford measurements, we find that the magic saturates in both phases. Our work sheds light on the behavior of magic and its linear combinations in quantum circuits, employing genuine magic measures. 
\end{abstract}

\maketitle

\section{Introduction}
The resources of quantum physics allow us to reach an advantage over classical simulations. It is known that entanglement represents a necessary resource to achieve this goal, and has been thoroughly studied~\cite{amico2008entanglement,eisert2010colloquium,Horodecki2009}. However, probing entanglement is insufficient for quantum advantage~\cite{vidal2003efficient}. For instance, there is a class of states called the stabilizer states~\cite{nielsen2002quantum} that can be efficiently simulated classically in polynomial time~\cite{gottesman1997stabilizer,gottesman1998theory,aaronson2004improved,veitch2014resource}. Stabilizer states are constructed by Clifford circuits, which are generated by the non-universal gate set of Hadamard, $\pi/4$ phase, and controlled-not gates. Therefore, any quantum circuit that should achieve a quantum advantage must include not only high amount of entanglement but also non-Clifford resources. Any extension of Clifford circuits enables them to perform universal quantum computation by allowing the input states to include the so-called magic states \cite{bravyi2005UniversalQuantumComputation,campbell2017roads}, i.e., states that are not stabilizer states. The canonical magic-state is $|T\rangle=\left( |0\rangle +e^{i\pi/4} |1\rangle \right)/ \sqrt{2}$ which enables the application of a single-qubit unitary $T=\rm{diag}(1,e^{i\pi/4})$, which, combined with the Clifford gates, creates a universal gate set \cite{nielsen2002quantum}. The amount of non-Clifford resources necessary to prepare a state is called nonstabilizerness, commonly referred to as ``magic'' \cite{bravyi2005UniversalQuantumComputation,campbell2017roads,bravyi2012magic,harrow2017quantum}.  

The importance of entanglement has stimulated interest in different fields. A prime example of this is hybrid quantum circuits that consist of unitary gates interspersed with measurements~\cite{potter_entanglement_2022, skinner_measurement-induced_2019, li_quantum_2018, li_measurement-driven_2019}.  These circuits can exhibit so-called entanglement transitions, where the system undergoes a shift between phases characterized by volume-law and area-law entanglement scaling. Numerical studies have extensively explored these transitions in various settings, such as random Clifford \cite{li_quantum_2018,li_measurement-driven_2019} or Haar circuits \cite{skinner_measurement-induced_2019} combined with projective measurements. Subsequent research has established that even circuits solely comprised of non-commuting measurements can exhibit entanglement transitions \cite{Ippoliti2021,Sang2021,Lang2020,Lavasani2021,Lavasani2021-2}. 

Despite its crucial role in achieving quantum advantage, little is known about the analogous magic transitions in monitored random quantum circuits. Recent studies have provided evidence for the existence of magic transitions in different contexts~\cite{leone_phase_2023,niroula_phase_2023,bejan2023dynamical,fux2023entanglement}. In particular, Refs \cite{bejan2023dynamical, fux2023entanglement} investigated magic transitions in the context of monitored Clifford circuits doped by $T$ gates. As mentioned above, Clifford gates along with the $T$ gate form a universal gate set for quantum computation. Thus, such $T$-doped Clifford circuits interpolate between classically simulable and universal circuits, and the above works found that the two limits are separated by a transition in magic. In Ref. \cite{bejan2023dynamical}, Bejan et al computed the magic using Pauli-based computation \cite{Bravyi2016trading} that essentially maps the quantum dynamics to a magic state register subject to mutually commuting measurements. They found cases where magic and entanglement transitions coincide, but also others with a magic transition in a volume-law entangled phase. Instead in Ref. \cite{fux2023entanglement}, Fux et al studied both magic and entanglement transition using matrix product states (MPS) simulations, providing evidence that a transition in magic can occur independently of one in entanglement. However, both studies have some limitations. Ref. \cite{bejan2023dynamical} only computed a proxy of magic which can increase under Clifford operations, while the results presented in \cite{fux2023entanglement} may suffer systematic errors due to MPS truncation. As such, a proper characterization of magic transition using a true measure of magic remains an outstanding challenge.

Investigating magic transitions in quantum circuits presents significant challenges compared to entanglement transitions. While large-scale simulations of entanglement transitions often rely on efficiently simulable Clifford circuits, these circuits are inherently incapable of hosting magic transitions. In this paper, we introduce and study the magic in a measurement-only circuit consisting of Clifford and non-Clifford measurements, depicted in Fig. \ref{fig:PMcircuit}. Here, differently from previous studies, magic is injected through the non-Clifford measurements. We show that the magic dynamics in this circuit is efficiently simulable, employing any measure of magic that is additive for all tensor products of single-qubit states. This allows us to perform large-scale simulations to study the magic transition in this circuit, which can be viewed as a result of the competition between Clifford and non-Clifford measurements. 

The study of entanglement transitions has benefited significantly from the construction of linear combinations of entanglement measures \cite{Zabalo2020,Lavasani2021,Lavasani2021-2,Klocke2022,Kells2023}. Motivated by this success, we initiate a parallel investigation in the context of magic. We analyze the mutual magic and topological magic, which will be defined according to some partitioning of the system (see Fig. \ref{fig:fig2}(b)). Quantifying magic in mixed states, which arise when considering subsystems, is notoriously difficult compared to pure states. For qubits, there are currently no known computable measures for mixed states. Despite the general difficulty, we can demonstrate that in our specific setup, the magic of subsystems exhibits a simplified form, allowing us to leverage existing, robust measures of magic.

With constant density of non-Clifford measurements per time step, our results demonstrate that the magic scales extensively with no distinctive features near the percolation critical point. However, mutual magic serves as a clear indicator of the transition, showcasing a distinct peak at the critical point. Specifically, it displays a logarithmic divergence with system size in one dimension and an area-law scaling in two dimension, analogous to entanglement entropy. Further analysis using topological magic enables precise finite-size scaling, allowing us to extract critical exponents which are found to match the bond percolation values. On the other hand, with vanishing non-Clifford measurement rate, we found that the magic saturates to a constant, in agreement with previous studies. Finally, we discuss a specific scenario where the dynamics of mutual magic is exactly identical to the entanglement dynamics. Overall, our work provides a genuine understanding of the nontrivial dynamics of magic and its linear combinations, which importantly utilizes true measures of magic.

The rest of the paper is structured as follows. In Sec. \ref{sec:model}, we introduce the quantum circuit with Clifford and non-Clifford measurements. In Sec. \ref{sec:simul}, we present a classical simulation of the circuit. 
In Sec. \ref{sec:measures}, we discuss the magic properties of the circuit and introduce mutual magic and topological magic. In Sec. \ref{sec:numerics}, we present our numerical results in both one- and two-dimensional lattices. In Sec. \ref{sec:pce}, we briefly comment on the connection to the participation entropy and then conclude in Sec. \ref{sec:conclusion}.

\section{Model} \label{sec:model}
\begin{figure}
\centering
\includegraphics[width=.95\linewidth]{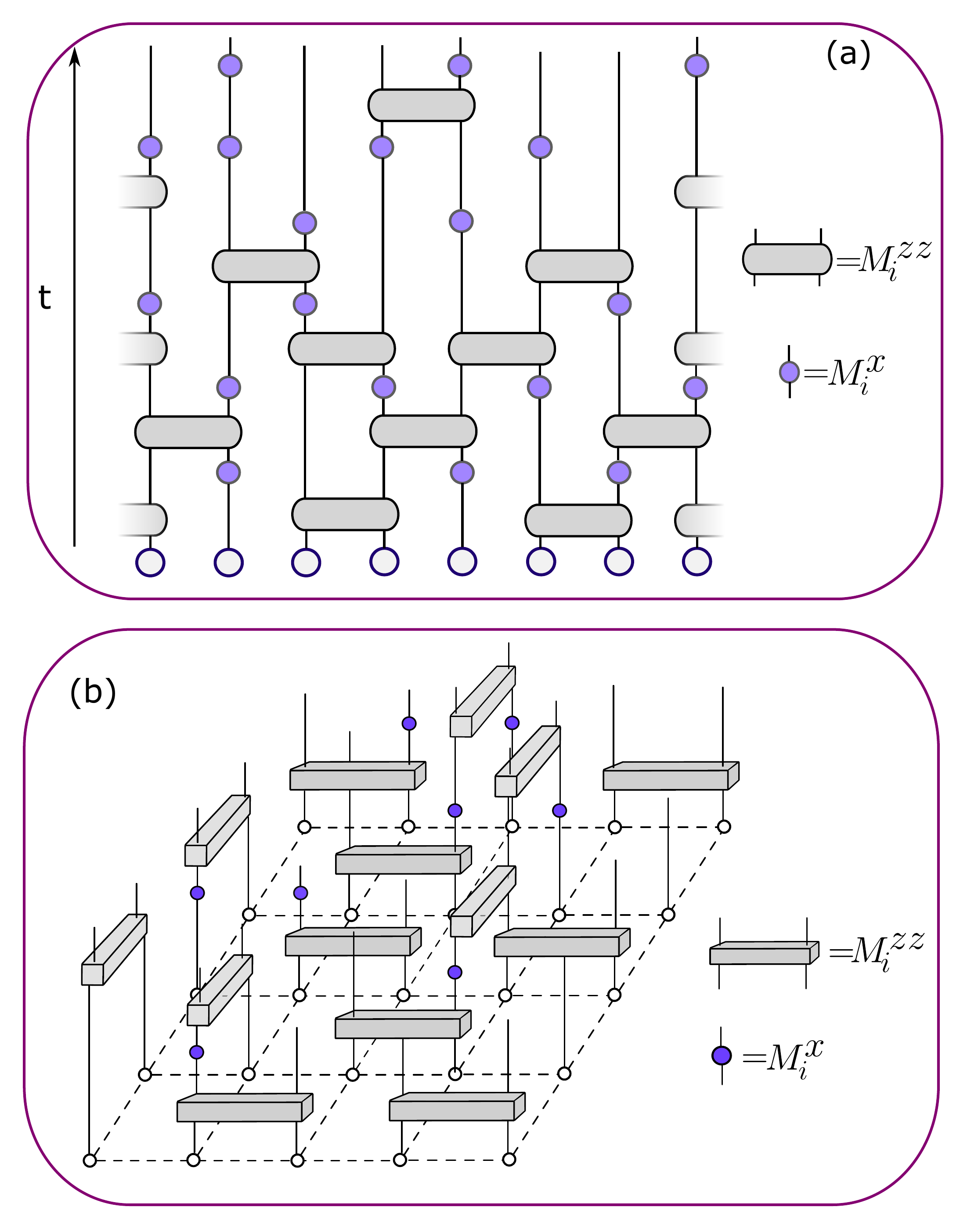}
\caption{Measurement-only quantum circuit with two types of competing measurements in (a) one-dimensional and (b) two-dimensional lattices. Gray boxes on edges denote measurements $M_{zz}$ on adjacent spins and violet circles denote measurements $M_x$ on a single spin. Each time step comprises one row of $M_{zz}$ measurements followed by a row of $M_x$ measurements.}
\label{fig:PMcircuit}
\end{figure}

Consider a system with spin-1/2 degrees of freedom in every site $i$. Each spin is represented by Pauli matrices $\sigma^{\alpha}_i$ with $\alpha=\left\lbrace x,y,z \right \rbrace$. The quantum circuit is defined by projective measurements of observables $O$, and the action of such a measurement is given by 
\be 
M[O] |\psi \rangle = \frac{P_{\lambda} |\psi \rangle}{\sqrt{\langle \psi | P_{\lambda} |\psi \rangle }},
\ee
which is the post-measurement state after measurement of the discrete eigenvalue $\lambda$ of $O$ with probability $\mathrm{Pr}(\lambda)=\langle \psi | P_{\lambda} |\psi \rangle$. Here, $P_{\lambda}$ denotes the projector onto the corresponding eigenspace. 
We are interested in measurements of the observables $\tilde{\sigma}^x_i(\theta)=e^{-i\theta/2 \sigma^z_i} \sigma^x_i e^{i\theta/2 \sigma^z_i}$ and $\sigma^z_i \sigma^z_{i+1}$. The angle $\theta$, which can vary in space and time, will play an important role on the behavior of magic, as discussed further below. The eigenvectors of $\tilde{\sigma}^x_i(\theta)$ are $|\pm_{\theta}\rangle=|0\rangle \pm e^{i \theta} |1\rangle$. The projectors associated with the two measurements are 
\be \label{eq:projector_sigmax} 
P^{\tilde{\sigma}^x}_{\lambda}=\frac{1}{2} \left(1+\lambda \tilde{\sigma}^x \right)
\ee
\be \label{eq:projector_sigmaz}
P^{\sigma^z_i \sigma^z_{i+1}}_{\lambda}=\frac{1}{2} \left(1+\lambda \sigma^z_i \sigma^z_{i+1} \right),
\ee
with the set of outcomes $\lambda \in \{+1,-1 \}$. Each time step comprises one row of $M_{zz}$ measurements followed by a row of $M_x$ measurements. Each edge $e=(i,i+1)$ is measured by the observable $\sigma^z_i \sigma^z_{i+1}$ with probability $1-p$ and each site $i$ is measured by the observable $\tilde{\sigma}^x_i(\theta)$ with probability $p$ (see Fig. \ref{fig:PMcircuit}). Given the state $|\psi(t) \rangle$ at time $t$, the new wave function at $t + 1$ is given by
\be 
|\psi(t+1) \rangle = M_x M_{zz} |\psi(t) \rangle 
 \ee
with measurements
\be 
M_{x}= \prod_i  M^x_i \quad M_{zz}= \prod_i  M^{zz}_i.
\ee 
For any realization of such a circuit there is an ensemble of quantum trajectories of pure states, where each trajectory is labeled by the sequence of measurement outcomes. We are interested in the long time limit of magic averaged over both quantum circuit realizations and quantum trajectories.

\section{Classical simulation} \label{sec:simul}
\begin{figure}
\centering
\includegraphics[width=.95\linewidth]{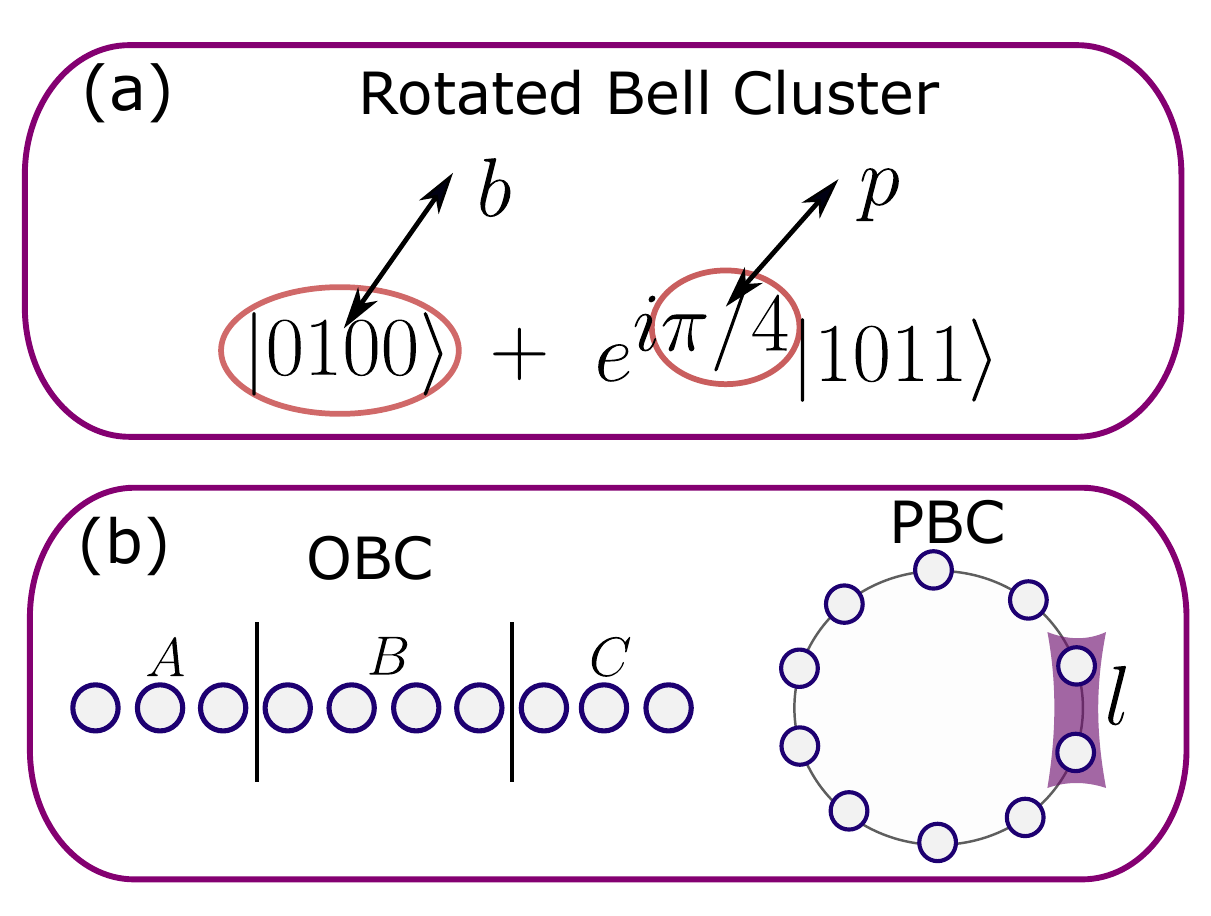}
\caption{(a) Sketch of rotated Bell cluster. (b)  Schematics of partitions: in the left part we show the partition for an open chain for the calculation of topological magic in Eq. \eqref{eq:topo_magic}. In the right part we show the partition for periodic boundary condition for the calculation of mutual magic in Eq. \eqref{eq:mutual_magic}.}
\label{fig:fig2}
\end{figure}

For the circuit described above, it can be shown that the state of the system can always be described as a tensor product of ``rotated Bell clusters'' (RBCs) defined as states that can be written as 
\begin{equation}
    \Ket{\mathbf{m}}  + e^{i \theta}\Ket{\overline{\mathbf{m}}},
\end{equation}
where $\overline{\mathbf{m}} = \prod_i \sigma^x_i \mathbf{m}$, as depicted in Fig. \ref{fig:fig2}(a). To see this, let us consider the example of two-qubit:
\begin{itemize}
    \item 
        Measuring $\Tilde{\sigma}^x_1(\theta)$ in the two-qubit RBC
        \begin{equation}
            \begin{split}
                &\Ket{00}+e^{i \varphi}\Ket{11}\\
                =&(\Ket{+_{\theta}0}+e^{i (\varphi-\theta)}\Ket{+_{\theta}1})+(\Ket{-_{\theta}0}-e^{i (\varphi - \theta)}\Ket{-_{\theta}1})
            \end{split}
            \label{eq:ex3}
        \end{equation}
        yields either $\Ket{+_{\theta}}\otimes(\Ket{0}+e^{i (\varphi-\theta)}\Ket{1})$ or 
        $\Ket{-_{\theta}}\otimes(\Ket{0}-e^{i (\varphi-\theta)}\Ket{1})$ with equal probability. Notice that all the states are RBCs. The number of clusters in the state is increased by one.
    \item
        Measuring $\sigma_1^z\sigma_2^z$ in the product state
        \begin{equation}
            \begin{split}
                &(\Ket{0}+e^{i \varphi_1}\Ket{1})\otimes(\Ket{0}+e^{i \varphi_2}\Ket{1})\\
                =&(\Ket{00}+e^{i (\varphi_1+\varphi_2)}\Ket{11})+(e^{i \varphi_2}\Ket{01}+e^{i \varphi_1}\Ket{10})
            \end{split}
            \label{eq:ex2}
        \end{equation}
        yields either $\Ket{00}+e^{i (\varphi_1+\varphi_2)}\Ket{11}$ or
        $\Ket{01}+e^{i (\varphi_1 - \varphi_2)}\Ket{10}$ with equal probability. 
        Notice that all the states are again RBCs. This process can be seen as a merging of RBCs.
\end{itemize}
Generalization to higher number of qubits is straightforward.

        Consequently, the circuit can be efficiently simulated by a classical stochastic
process. The state of the system is characterized by vectors $\mathbf{s}\in\mathbb{N}_0^{L}$ and $\mathbf{b}\in\mathbb{Z}_2^{L}$. The nonnegative
        integer $s_i\in\mathbb{N}_0$ encodes that site $i$ belongs to an RBC with label $s_i$. Moreover, an RBC labeled by $n$ is associated with a phase $p_n$. Let $A_n = \{q_i\}$ be the set of sites that belongs to the RBC $n$. 
        The state of the RBC reads
        \begin{equation}
            \Ket{\mathbf{b}_i}_{i \in A_n}  + e^{i p_n}\Ket{\overline{\mathbf{b}}_i}_{i \in A_n}
        \end{equation}
        Since the state is simply a product states of RBCs, the vectors $\mathbf{s}\in\mathbb{N}_0^{L}$, $\mathbf{b}\in\mathbb{Z}_2^{L}$ and the phase $p_n$ for each RBC completely specify the state. Moreover, they can be updated very efficiently, as we shall discuss in detail below.

        We will provide the update rule for the two types of measurements. Here, sites that are not mentioned remain unchanged.
        \begin{itemize}
            \item \textit{Measurement of $\Tilde{\sigma}^x_i(\theta)$.}
            The outcome is $\lambda=\pm 1$ with equal probability. Set $s_i':=\operatorname{next}(\mathbf{s})$. Here,
                $\operatorname{next}(\mathbf{s})=\min(n\in\mathbb{N}_0\setminus
                \mathbf{s})$ returns the smallest integer that is not currently used as a cluster label in $\mathbf{s}$. Set $b_i'=0$, $p_{s_i}'=p_{s_i'} - (-1)^{b_i} \theta + \delta_{\lambda,-1} \pi$, and $p_{s_i'}'=\theta + \delta_{\lambda,-1} \pi$.
            \item \textit{Measurement of $\sigma^z_i \sigma^z_{j}$.} The outcome is $\lambda=\pm 1$ with equal probability. Set $s_l'=s_i$ for all sites $l$ with $s_l=s_j$. There are two cases:
            \begin{itemize}
                \item $\lambda=1-2 b_i \oplus b_j $. Set $p_{s_i}'=p_{s_i}+p_{s_j}$. 
                \item $\lambda=2 b_i \oplus b_j -1$.
                Set $b_l'=1-b_l$ for all sites $l$ with $s_l=s_j$. Set $p_{s_i}'=p_{s_i}-p_{s_j}$.
            \end{itemize}
        \end{itemize}

        Notice that the dynamics of the the vectors $\mathbf{s}$ and $\mathbf{b}$ are not affected by the angles $\theta$. Thus, they are identical also to the stabilizer case when all $\theta=0$, whose entanglement transition has been studied in Ref. \cite{Lang2020} (see also \cite{Nahum2020}). Setting nonzero $\theta$ is however crucial to induce nontrivial magic dynamics. In the following, we perform the simulation of the circuit using the update rules above. We have also benchmarked our results numerically against MPS simulations \cite{haug2023quantifying,Tarabunga2024}. 

\section{Magic measures} \label{sec:measures}

The fact that the states at each time step are composed of RBCs offers a significant advantage for quantifying magic. To see this, consider an RBC with size $L_B$. We can define a Clifford unitary consisting of CNOT operations $C=CNOT_{1,2} CNOT_{2,3} \dots CNOT_{L_B-1,L_B}$. Applying $C$ to the RBC $\ket{\psi} = \Ket{\mathbf{b}}  + e^{i \theta}\Ket{\overline{\mathbf{b}}}$ results in the state $C \ket{\psi} = (\ket{b_0} + e^{i \theta} \ket{\overline{b_0}}) \otimes \ket{c_2} \otimes \dots \otimes \ket{c_{L_B}} $, where $c_k = b_{k-1} \oplus b_k$. In other words, RBCs can be transformed to a product state of a single-qubit magic state and stabilizer states by applying Clifford unitaries. This observation is crucial because magic measures must be invariant under Clifford unitaries and composition with stabilizer states \cite{veitch2014resource}. It follows that the magic within the original circuit can be determined solely by considering the tensor product of these single-qubit magic states. This significantly simplifies the task of calculating magic in such circuits.

Our analysis extends to investigating the magic of subsystems $\rho_A = \Tr_{A^c}[|\psi \rangle \langle \psi |]$, where $A^c$ is the complement of $A$. The key point is that partially tracing an RBC yields a classical mixture of two computational basis states, irrespective of its phase. Therefore, the reduced density matrix has the form of a tensor product of pure RBCs and classical mixtures. Since such a classical mixture is a mixed stabilizer state, the magic of $\rho_A$ can again be reduced to the magic of tensor-product of single-qubit pure magic states.

The above observations pave the way for efficient magic calculations in these circuits. Since the magic of the entire system boils down to the magic of single-qubit pure magic states in a tensor product structure, we can leverage magic measures that exhibit a key property: additivity for all tensor products of single-qubit states. Such measures exist \cite{Bravyi2019,Seddon2021,rubboli2023mixedstateadditivitypropertiesmagic,leone2022stabilizer,Beverland2020}, including those whose original definitions involve minimization procedures, and are thus generally difficult to compute beyond a few qubits. Importantly, this includes \textit{bona fide}, strong measures of magic for both pure states and mixed states, such as the relative entropy of magic \cite{veitch2014resource}. Hereafter, we will use $\mathcal{M}$ to denote any measure of magic that is additive for all tensor products of single-qubit states. For any $\mathcal{M}$, the magic of the full state is simply given by the total sum of the magic of individual RBCs, measured by $\mathcal{M}$.

For a magic measure $\mathcal{M}$, we will consider the ``mutual magic'', defined in a subsystem $A$ as
\begin{equation} \label{eq:mutual_magic}
    I_\mathcal{M}(A) = \mathcal{M}(| \psi \rangle) - \mathcal{M}(\rho_{A}) -\mathcal{M}(\rho_{A^c}).
\end{equation}
We will use the notation $ I_\mathcal{M}(\ell) $ to denote the case $A=  \{ 1,...,\ell \}$. This quantity can be viewed as the amount of magic that resides in the correlations between subsystems. A similar quantity has been studied previously in the context of mana \cite{tarabunga2023critical,white2021} and stabilizer R\'enyi entropy \cite{tarabunga2023manybody,Frau2024,lópez2024exact}, where it was shown that such mutual-information-like quantity is able to detect the transition when the full-state magic does not show any features at the transition.

In terms of RBCs, $I_\mathcal{M}(A)$ is given by the sum of the magic of RBCs with support both in $A$ and $A^c$. This interpretation offers a physical picture of mutual magic as entanglement modulo stabilizer contributions. In particular, it immediately follows that it is upper bounded by the entanglement entropy \footnote{Here, we assume that $\mathcal{M}$ is upper bounded by one for single-qubit states, which is typically the case (or can be made so after rescaling).}:
\begin{equation}
    I_\mathcal{M} (A)  \leq S(A).
\end{equation}

Finally, in order to distinguish the magic content between different phases, we will consider the topological magic defined as 
\begin{equation} \label{eq:topo_magic}
    \mathcal{M}^t_{\mathrm{topo}} = \mathcal{M}(\rho_{ABC}) + \mathcal{M}(\rho_B) - \mathcal{M}(\rho_{AB}) - \mathcal{M}(\rho_{BC}),
\end{equation}
 for systems with open boundary condition. Here, the system is divided into three non-overlapping parts $A,B,$ and $C$. Such linear combination was first introduced in the context of entanglement by the name of ``generalized topological entanglement entropy'' \cite{zeng2015,Zeng2019} to probe symmetry-breaking orders. The latter has also been considered in the context of measurement-induced entanglement transition \cite{Lavasani2021,Klocke2022}. In our setup, $\mathcal{M}^t_{\mathrm{topo}}$ is given by the sum of the magic of RBCs with support in $A,B$ and $C$.

\section{Numerical results} \label{sec:numerics}
\subsection{Magic in (1+1)D circuits}
We will focus on the case of fixed $\theta=\pi/4$. In this case, the possible phases of the RBCs become restricted to multiples $\pi/4$. If the phase is a mutiple of $\pi/2$, the state is a stabilizer state. If it is not, then it is equivalent to the canonical $T$ state, up to a Clifford unitary. We will denote the magic of a $T$ state as $\mathcal{M}_T$. Applying the update rule in Sec. \ref{sec:simul}, one can see that any RBC of even size is a stabilizer state, while an RBC of odd size has the magic equal to $\mathcal{M}_T$. This simplification allows for faster simulations by solely tracking the parity of the sizes of the RBCs.

\begin{figure}
\centering
\includegraphics[width=.98\linewidth]{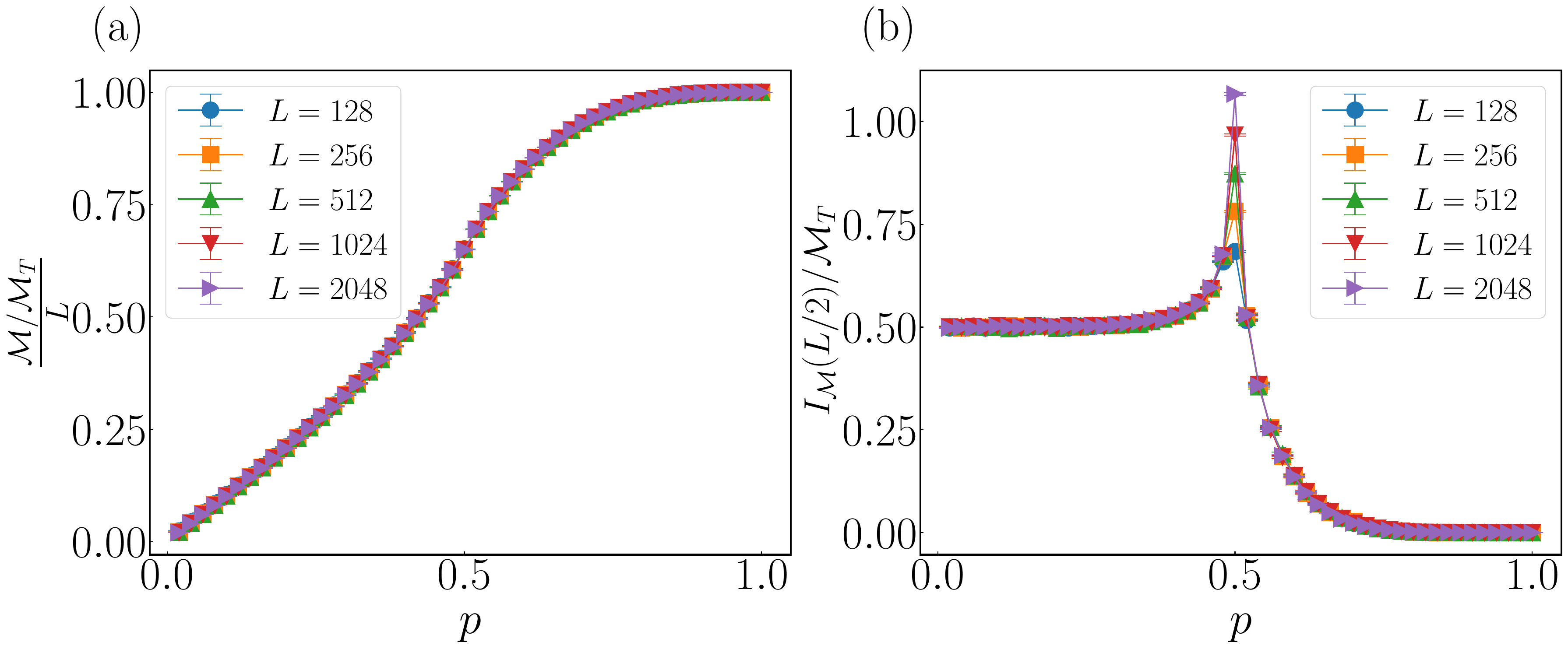}
\caption{ (a) Magic density $\frac{\mathcal{M}/\mathcal{M}_T}{L}$ and (b) mutual magic of half subsystem $I_\mathcal{M}(L/2)/\mathcal{M}_T$ with periodic boundary condition. }
\label{fig:nullity}
\end{figure}

\begin{figure}
\centering
\includegraphics[width=.98\linewidth]{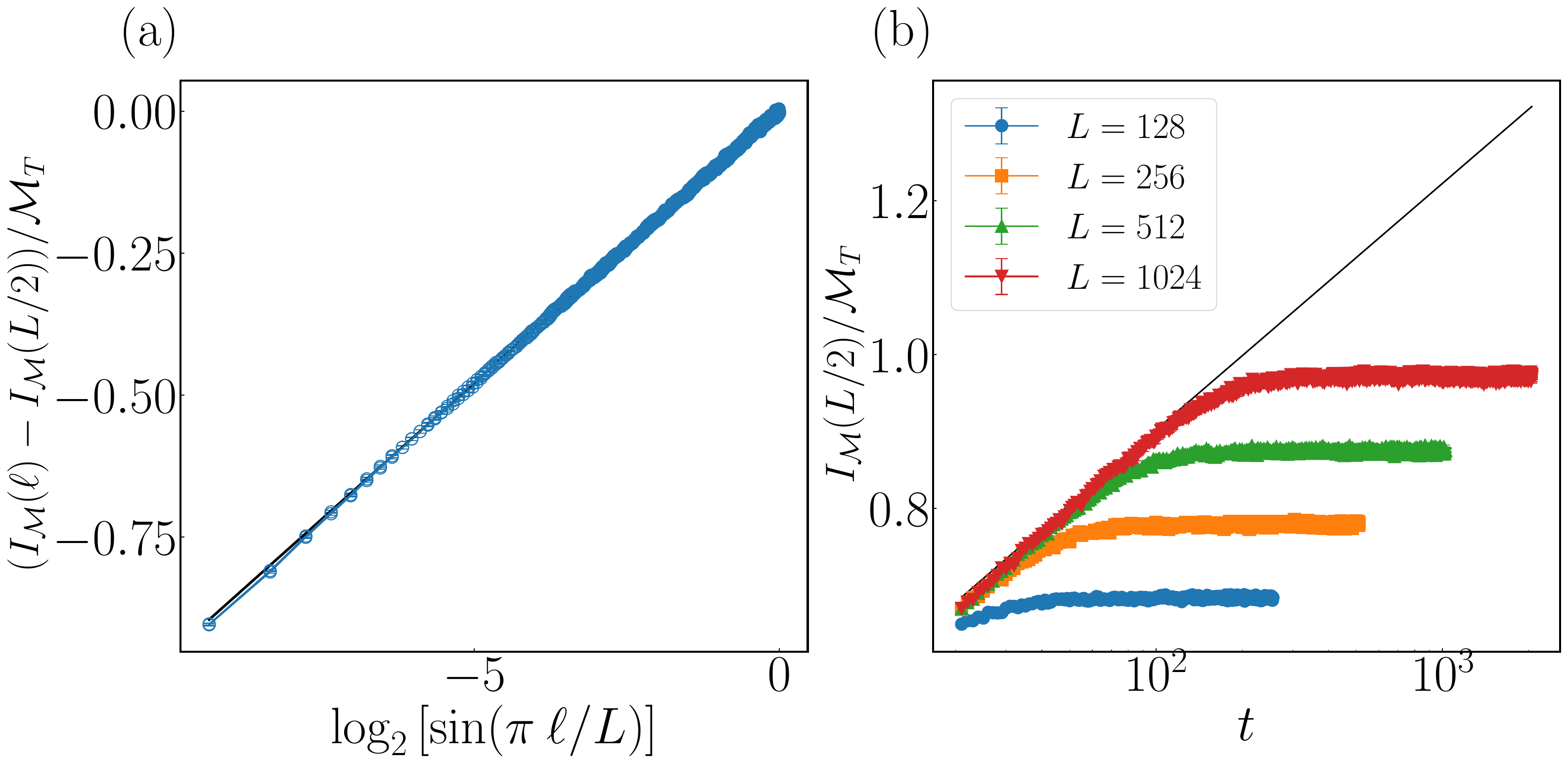}
\caption{(a) Mutual magic $(I_\mathcal{M}(\ell) - I_\mathcal{M}(L/2))/\mathcal{M}_T$ at $p=p_c$ and $L=2048$ and (b) the dynamics of $I_\mathcal{M}(L/2)/\mathcal{M}_T$ as a function of time $t$ for different system sizes with periodic boundary condition. The black line denotes a linear fit. }
\label{fig:nullity_pc}
\end{figure}

We now present our numerical results. We start with the initial product state $\ket{\psi(0)} = \ket{+_\theta}^{\otimes L}$ and run the circuit for $2L$ time steps for the system to reach the steady state. We then average the quantities of interest over $10^4 - 10^5$ trajectories.
We show the magic and the mutual magic in Fig. \ref{fig:nullity} as a function of $p$ for systems with periodic boundary condition. Here, we only consider even $L$, such that $\mathcal{M}$ vanishes for $p=0$ as the state is a global (stabilizer) Bell cluster. If $L$ is odd, the state becomes a global RBC, whose magic is $\mathcal{M}=\mathcal{M}_T$. For $p=1$, the state is a tensor product of $T$ states, such that the magic is given by $\mathcal{M}/L=\mathcal{M}_T$. Our numerical results show that the magic scales extensively at any nonzero $p$, as shown in Fig. \ref{fig:nullity}(a). However, around the percolation transition at $p_c=0.5$ \cite{Sykes1963}, the magic appears rather featureless. Instead, the transition is clearly identified using the mutual magic, which appears to diverge with $L$ (see Fig. \ref{fig:nullity}(b)). This behavior is reminiscent of entanglement entropy, which grows logarithmically at $p_c=0.5$ \cite{Lang2020,Calabrese2004,Calabrese2009}:
\begin{equation}
    S(\ell) = \frac{\Tilde{c}}{3} \log_2 \left[ \frac{L}{\pi} \sin \left(\ell \frac{\pi}{L} \right) \right] + \gamma,
\end{equation}
where $\Tilde{c}=3 \sqrt{3} \ln(2)/(2 \pi) \approx 0.573$,  and $\gamma$ is a non-universal constant. We thus postulate that the mutual magic follows similar scaling:
\begin{equation} \label{eq:mutual_nullity_scaling}
    I_\mathcal{M}(\ell) = \frac{\Tilde{c}_\mathcal{M}}{3} \log_2 \left[ \frac{L}{\pi} \sin \left(\ell \frac{\pi}{L} \right) \right] + \gamma'.
\end{equation}
We show the scaling of $I_\mathcal{M}(\ell) - I_\mathcal{M}(L/2)$ at $p_c$ in Fig. \ref{fig:nullity_pc}(a), which confirms the hypothesis in Eq. \eqref{eq:mutual_nullity_scaling}. Indeed, we observe that $I_\mathcal{M}(\ell) - I_\mathcal{M}(L/2) \approx \frac{\Tilde{c}_\mathcal{M}}{3} \log_2 \left[  \sin \left(\ell \frac{\pi}{L} \right) \right]$, where $\Tilde{c}_\mathcal{M} \approx  \mathcal{M}_T  \Tilde{c} /2$.

We further investigate the magic growth with time, particularly at the critical point. At criticality, and for timescales significantly shorter than the saturation time, we postulate,
\begin{equation} \label{eq:mutual_nullity_dynamics}
    I_\mathcal{M}(\ell, t) =  \frac{\Tilde{c}_{\mathcal{M},t}}{3} \log_2 t + \gamma''
\end{equation}
where $\gamma''$ is a non-universal constant. Fig. \ref{fig:nullity_pc}(b) shows the dynamics of $I_\mathcal{M}(L/2)$ as a function of time $t$, which supports the postulate in Eq. \eqref{eq:mutual_nullity_dynamics}. Remarkably, we find that $\Tilde{c}_{\mathcal{M},t}=\Tilde{c}_{\mathcal{M}}$, mirroring the behavior observed for entanglement in conformal field theories (CFTs) with dynamical exponent $z=1$.

\begin{figure}
\centering
\includegraphics[width=.98\linewidth]{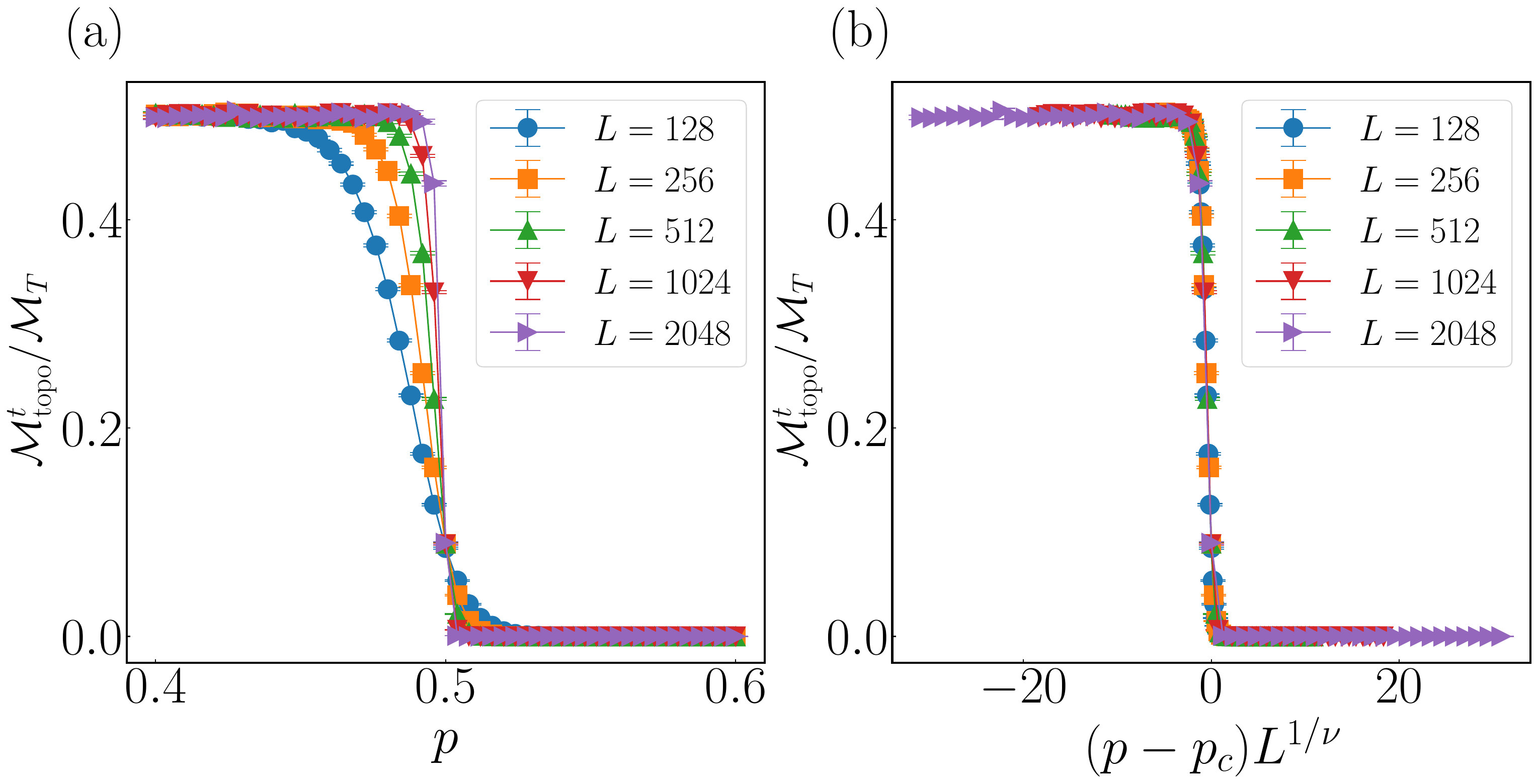}
\caption{ (a) The topological magic $\mathcal{M}^t_{\mathrm{topo}}/\mathcal{M}_T$ with open boundary condition. (b) Data collapse for the topological magic, showing excellent agreement with the percolation value $p_c=0.5$ and $\nu=4/3$. }
\label{fig:nullity_topo}
\end{figure}

We then turn to open boundary condition and investigate the topological magic $\mathcal{M}^t_{\mathrm{topo}}$. Here, we set $L=4L_A=4L_C=2L_B$. The results are shown in Fig. \ref{fig:nullity_topo}. It is clear that $\mathcal{M}^t_{\mathrm{topo}}$ tends to $0.5 (0)$ for $0<p<p_c (p>p_c)$ with increasing system sizes. This is again reminiscent of the behavior of the generalized topological entanglement entropy; however, the topological magic does not seem to be quantized to an integer value. Note that, for $p=0$, $\mathcal{M}^t_{\mathrm{topo}}$ vanishes since the magic itself vanishes (for even $L$). However, we found that $\mathcal{M}^t_{\mathrm{topo}} \to 0.5$ for any infinitesimal value of $p$. This observation can be explained as follows: for $p \to 0$, a macroscopic Bell cluster emerges \cite{Lang2020}, which weight have the same probability of being odd or even, on general grounds.

We further investigate the finite-size scaling of $\mathcal{M}^t_{\mathrm{topo}}$, using the finite-size scaling hypothesis:
\begin{equation}
    \mathcal{M}^t_{\mathrm{topo}} = f \left( L^{1/\nu}(p-p_c) \right),
    \label{eq:fss}
\end{equation}
where $f(x)$ is some unknown function, $p_c$ is the critical value of tuning parameter $p$, and $\nu$ is the correlation length critical exponent. We found that our numerical data exhibits excellent data collapse with $p_c=0.5$ and $\nu = 4/3$, expected from 2D bond percolation.

\begin{figure}
\centering
\includegraphics[width=.98\linewidth]{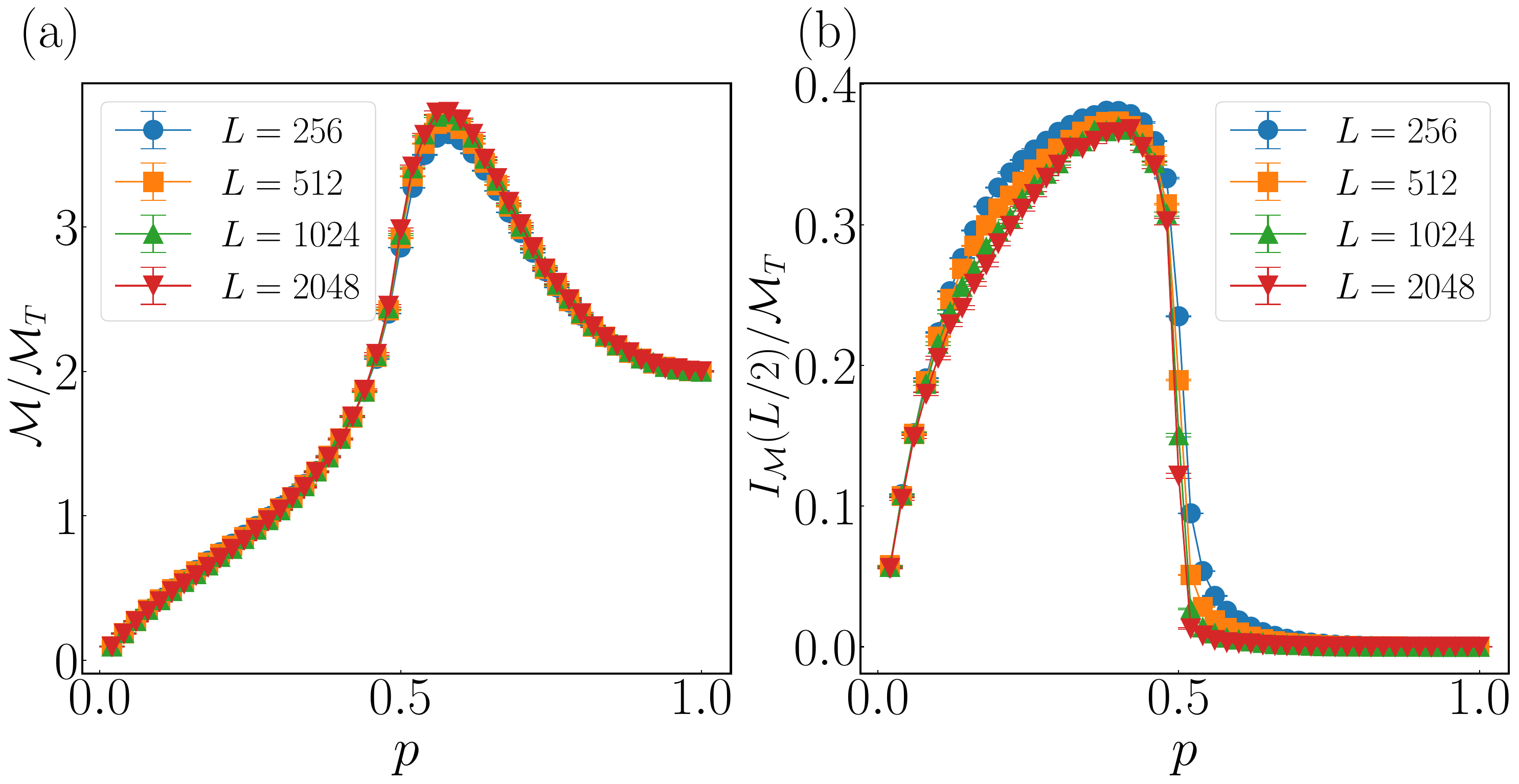}
\caption{ (a) Magic density $\frac{\mathcal{M}/\mathcal{M}_T}{L}$ and (b) mutual magic of half subsystem $I_\mathcal{M}(L/2)/\mathcal{M}_T$ with periodic boundary condition and $q=2/L$. }
\label{fig:nullity_2L}
\end{figure}

Let us now consider a different scenario. We introduce a parameter $q$, such that the measurement of $\Tilde{\sigma}^x$ is performed at an angle $\theta=\pi/4$ with probability $q$ and $\theta=0$ (Clifford measurement) with probability $1-q$. Note that the previous scenario corresponds to $q=1$. To mimic previous studies \cite{fux2023entanglement,bejan2023dynamical} where non-Clifford operations occur at a vanishing rate, we set $q=2/L$. The magic behavior is shown in Fig. \ref{fig:nullity_2L}. In contrast to the previous case, the magic no longer exhibits extensive scaling with system size. Instead, it saturates to a constant value. This observation confirms the existence of such $O(1)$ magic phase that emerges when non-Clifford operations are introduced at a vanishing rate. In this case, the mutual magic appears to play the role of an order parameter, being zero for $p>p_c$ and nonzero for $p<p_c$.

\subsection{Magic in (2+1)D circuits}
\begin{figure}
\centering
\includegraphics[width=.98\linewidth]{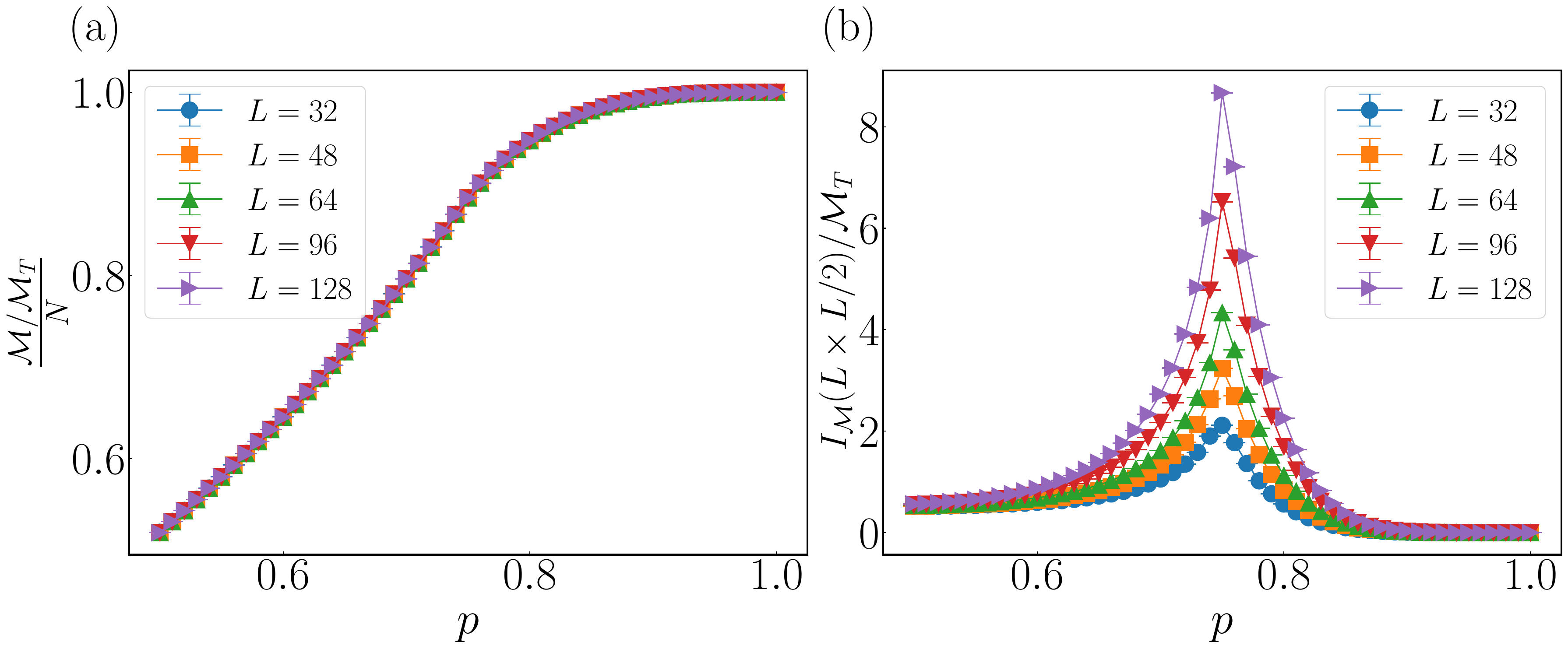}
\caption{ (a) Magic density $\frac{\mathcal{M}/\mathcal{M}_T}{L}$ and (b) mutual magic of half subsystem $I_\mathcal{M}(L \times L/2)/\mathcal{M}_T$ on 2D square lattice with periodic boundary conditions. }
\label{fig:nullity_2d}
\end{figure}

\begin{figure}
\centering
\includegraphics[width=.48\linewidth]{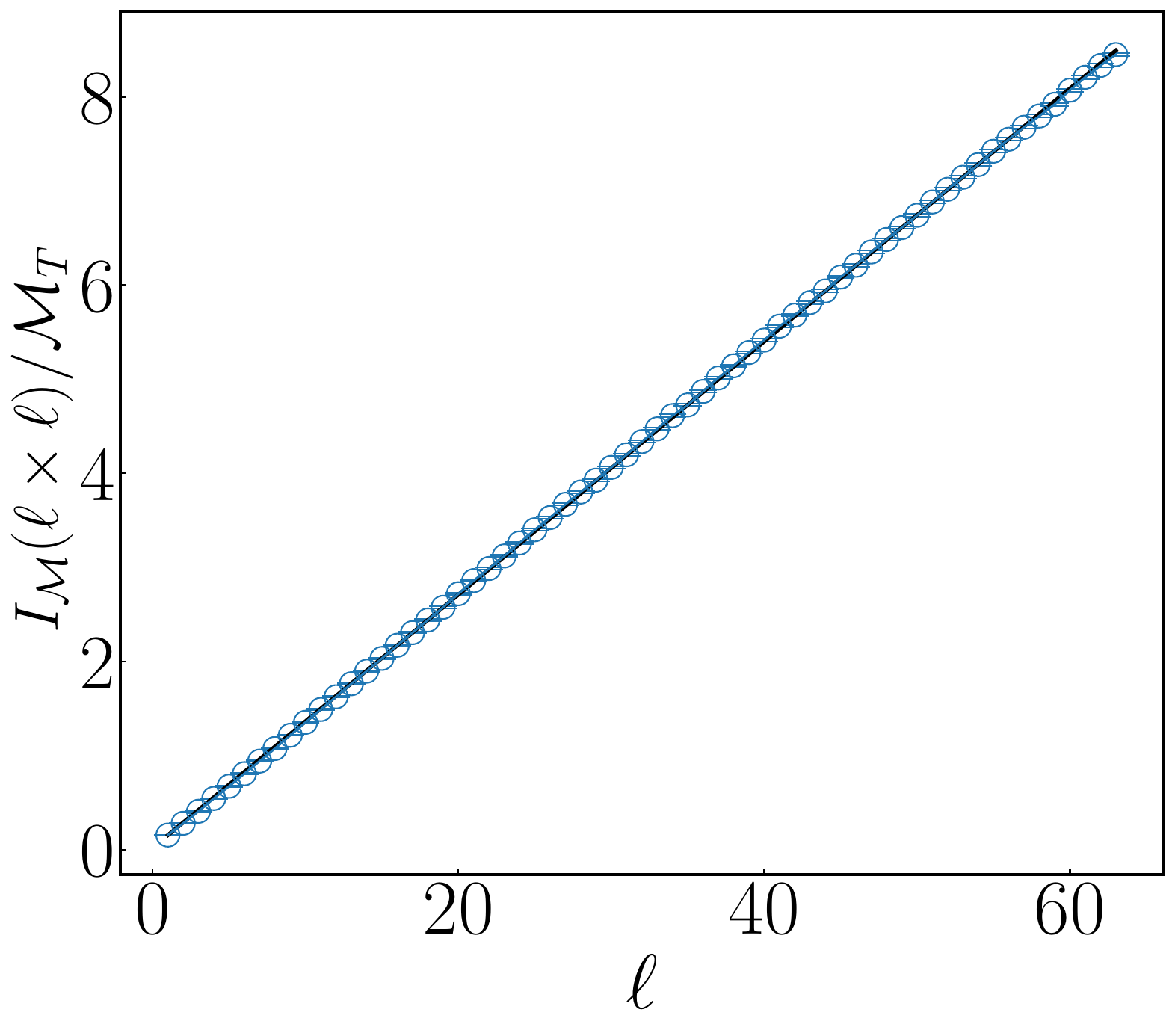}
\caption{Mutual magic $I_\mathcal{M}(\ell \times \ell)/\mathcal{M}_T$ at $p=p_c$ and $L=128$ on 2D square lattice with periodic boundary conditions. The black line denotes a linear fit. }
\label{fig:nullity_pc_2d}
\end{figure}

We extend our analysis to a 2D square lattice, where the model exhibits a connection to three-dimensional bond percolation on the cubic lattice with a critical rate $p_c \approx 0.75$, as numerically determined in Ref. \cite{vanderMarck1997}. We will again consider the scenario of fixed $\theta=\pi/4$. Employing a simulation procedure similar to the 1D case, we simulate systems of size $N = L \times L$ with periodic boundary conditions, and we run the circuit for $L$ time steps to reach the steady state. Fig. \ref{fig:nullity_2d} displays the behavior of magic and mutual magic ($I_\mathcal{M}(\ell_x \times \ell_y)$ for a region of size $\ell_x \times \ell_y$) as a function of the parameter $p$. Similar to the 1D scenario, magic remains featureless while mutual magic captures the transition at $p=p_c$. However, in the 2D case, the mutual magic exhibits area-law scaling at the critical point, as shown in Fig. \ref{fig:nullity_pc_2d}. This is again consistent with the behavior of entanglement.

\begin{figure}
\centering
\includegraphics[width=.98\linewidth]{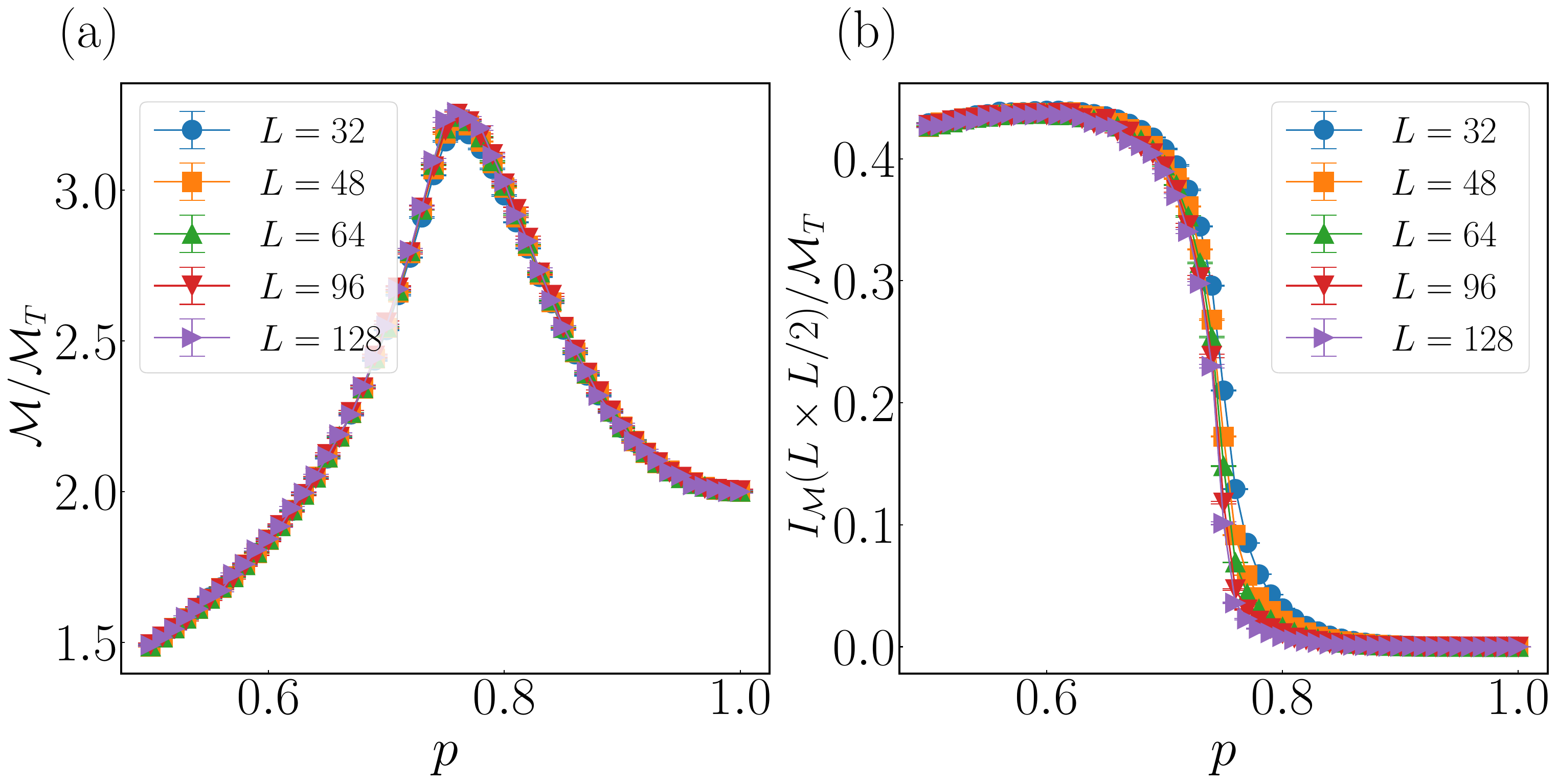}
\caption{ (a) Magic density $\frac{\mathcal{M}/\mathcal{M}_T}{L}$ and (b) mutual magic of half subsystem $I_\mathcal{M}(L/2)/\mathcal{M}_T$ on 2D square lattice with periodic boundary conditions and $q=2/N$. }
\label{fig:nullity_2d_2L}
\end{figure}

Setting $q=2/N$, we again observe a similar behavior as in the 1D case. The magic and mutual magic are shown in Fig. \ref{fig:nullity_2d_2L}. While the total magic saturates to a constant value, the mutual magic emerges as a clear order parameter, signaling the phase transition.

\subsection{Random $\theta$} \label{sec:random_theta}
We now discuss the case when the angles $\theta$ are chosen uniformly at random in the interval $[0,2\pi)$, both in space and time. We will show that the dynamics of mutual magic can be tuned to be identical to the entanglement.

To this end, we will focus on a specific measure of magic called the stabilizer nullity \cite{Beverland2020}. It is simply related to the size of the stabilizer group ${\rm Stab}(\psi)$, which is the group of Pauli strings that stabilize $| \psi \rangle$. The stabilizer nullity is defined as \cite{Beverland2020} 
\begin{equation}
    \nu(| \psi \rangle) = N - \log_2 \left( |{\rm Stab}(\psi)| \right).
\end{equation}
 It is known that $\nu$ is a strong magic monotone, which is also additive under tensor product. 

 Stabilizer nullity has only been formally defined for pure states. To analyze the mutual magic, we would need to extend it to mixed states. One possible extension is by using the convex roof construction:
 \begin{equation} 
    \nu(\rho) = \min_{\{p_i, \ket{\psi_i} \}}  \sum_i p_i  \nu(\ket{\psi_i}) ,
\end{equation}
where the minimum is taken over all possible convex pure-state decompositions of $\rho$: $\rho = \sum_i p_i \ket{\psi_i} \bra{\psi_i}$. Note that, the particular extension is not relevant, as long as it is invariant under composition with stabilizer states (see Sec. \ref{sec:measures}). The convex roof construction is convenient as it satisfies such condition.

A peculiar property of stabilizer nullity is that it can only take integer values. In particular, for a single-qubit state, it is zero for the single-qubit stabilizer states, and one otherwise. In terms of the RBCs, it is zero for an RBC with phase $\theta=k \pi/2$ with integer $k$, and one otherwise. With randomly chosen $\theta$, the probability of encountering an RBC as a stabilizer state is essentially zero. This implies that any RBC that has support in both $A$ and $A^c$ will contribute one to the mutual nullity $I_\nu(A)$, that is exactly the same procedure to compute entanglement entropy \cite{Lang2020}. We thus conclude that the mutual nullity in this setup is identical to the entanglement, as claimed.

\section{Connection to participation entropy} \label{sec:pce}
While this work focuses on the magic transition, we find it insightful to explore the connection with participation (Shannon) entropy, defined as
\begin{equation}
    S^{\mathrm{part}} (| \psi \rangle) =  \sum_\sigma  -\vert \langle \sigma | \psi \rangle\vert ^{2}\log_2  \vert \langle \sigma | \psi \rangle\vert^{2}.
\end{equation}
This quantity measures the spread of the wavefunction across different computational basis states. In our setup, one can see that it simply counts the total number of RBCs. This observation allows us to establish an inequality (that holds in our circuit):
\begin{equation}
    \mathcal{M}   \leq S^{\mathrm{part}}(\ket{\psi}).
\end{equation}
Therefore, like magic, the participation entropy is extensive at any nonzero $p$ \cite{Sierant2022}. Note that the inequality is saturated in the case of random $\theta$ (see Sec. \ref{sec:random_theta}).

For reduced density matrix $\rho_{A}$, the participation entropy is defined as
\begin{equation}
    S^{\mathrm{part}} (\rho_{A}) =  \sum_{\sigma_A}  - \langle \sigma_A \vert \rho_{A} \vert \sigma_A\rangle\log_2  \langle \sigma_A \vert \rho_{A} \vert \sigma_A\rangle,
\end{equation}
i.e., it is the Shannon entropy of the diagonal elements of $\rho_{A}$.
We can consider the Shannon mutual information \cite{alcaraz2013}:
\begin{equation} \label{eq:shannon_mutual}
    I_S(A) =  S^{\mathrm{part}}(\rho_{A}) +S^{\mathrm{part}}(\rho_{A^c}) - S^{\mathrm{part}}(| \psi \rangle).
\end{equation}
Previous studies have shown that it exhibits a scaling behavior similar to entanglement entropy in (1+1)D CFT \cite{alcaraz2013,stephan2014,alcaraz2015,alcaraz2015_2,alcaraz2016}. In our specific case, it is straightforward to see that the Shannon mutual information is exactly equal to the entanglement. This model thus provides an interesting example where the entanglement and Shannon mutual information exhibit the same scaling behavior, which can be understood on a microscopic level.

While the dynamics of entanglement and Shannon mutual information are not affected by the angles of $\Tilde{\sigma}^x$, we have seen in Sec. \ref{sec:numerics} that the dynamics of magic highly depends on them. By adjusting these angles (and potentially using different measures), we can manipulate the prefactor of the logarithmic scaling observed in the mutual magic, while the prefactors for entanglement and Shannon mutual information remain fixed. This highlights a significant distinction in how magic behaves compared to the other two resource quantities.

\section{Conclusions and outlook} \label{sec:conclusion}
In this work, we have introduced a measurement-only circuit which exhibits nontrivial magic dynamics. Notably, we show that, although the circuit is tuned away from the Clifford limit, it remains efficiently simulable through a mapping to a classical stochastic model. This allows for large-scale numerical simulations, revealing a competition between Clifford and non-Clifford measurements driving a magic transition between two distinct phases in which magic scales extensively with volume. These two phases can be distinguished from the topological magic, which takes a constant nonzero value for $p<p_c$, but vanishes for $p>p_c$. Furthermore, the mutual magic exhibits divergence at $p=p_c$, with logarithmic scaling in 1D and area-law scaling in 2D, similar to entanglement entropy. In the 1D case, this behavior aligns with the previous observation in (1+1)D CFT \cite{tarabunga2023critical}. Our results highlight the intriguing behavior of specific linear combinations of magic (that draws inspiration from entanglement), motivating further exploration in generic models. 

Interestingly, magic appears to exhibit distinct behavior compared to entanglement and participation entropy, as the latter two are completely independent of the angles of $\Tilde{\sigma}^x$. This raises general questions about the relationship between various quantum resources \cite{tirrito2023,turkeshi2023measuring,gu2024magicinducedcomputationalseparationentanglement} in a broader class of circuits.

Let us note that, unlike previous studies \cite{bejan2023dynamical,fux2023entanglement}, we find that the magic and entanglement transitions coincide. Future investigations could explore modifications, such as correlated monitoring \cite{bejan2023dynamical}, to potentially separate these transitions.
Moreover, our approach readily extends to simulating magic in symmetry-protected topological phases \cite{Lavasani2021} and toric code phases \cite{Lavasani2021-2}. This opens exciting avenues to explore the role of topology on magic dynamics. 

\acknowledgements
We thank M. Dalmonte, M. Frau, L. Piroli, T. Haug, R. Fazio, X. Turkeshi, G. Fux for collaboration on related topics. P. S. T. acknowledges support from the Simons Foundation through Award 284558FY19 to the ICTP.  ET was partially supported by QUANTERA DYNAMITE PCI2022-132919.

\bibliographystyle{apsrev4-1}
\bibliography{bibliography}

\end{document}